\documentclass[twoside,12pt]{article}
\usepackage{epsfig}

\newcommand{\be}{\begin{equation}}
\newcommand{\ee}{\end{equation}}
\newcommand{\bea}{\begin{eqnarray}}
\newcommand{\eea}{\end{eqnarray}}

\begin{document}

\title{Equation of state for dense supernova matter}

\author{Ch.C. Moustakidis\\
$^{}$ Department of Theoretical Physics, Aristotle University of
Thessaloniki, \\ 54124 Thessaloniki, Greece }

\maketitle

\begin{abstract}
We provide an equation of state for high density supernova matter
by applying a momentum-dependent effective interaction. We focus
on the study of the equation of state of high-density and
high-temperature nuclear matter containing leptons (electrons and
neutrinos) under the chemical equilibrium condition.  Thermal
effects on the properties and equation of state of nuclear matter
are evaluated and analyzed in the framework of the proposed
effective interaction model. Since supernova matter is
characterized by a constant entropy we also present  the
thermodynamic properties for the isentropic case
\cite{Moustakidis-09-b}.
\end{abstract}

\section{Introduction}

Knowledge of the properties of the equation of state (EOS) of hot
asymmetric nuclear matter is of fundamental importance to
understand the physical mechanism of the iron core collapse of a
massive star which produces a type-II supernova, and the rapid
cooling of a new born hot neutron star. Additionally, the EOS
defines the chemical composition, both qualitative and
quantitative, of the hot nuclear matter
\cite{Bethe-90,Prakash-97,Li-08,Bombaci-94}. Supernova matter
which exists in a collapsing supernova core and eventually forms a
hot neutron star at birth is another form of nuclear matter
distinguished in the participation of degenerate neutrinos and
electrons \cite{Takatsuka-94}. It is characterized by almost
constant entropy per baryon $S=1-2$ (in units of the Boltzmann
constant $k_B$) throughout the density $n$ and also by a high and
almost constant lepton fraction $Y_l=0.3-0.4$ in contrast with
ordinary neutron star matter where $S=0$ and $Y_l \leq 0.05$.
These characteristics are caused by the effects of
neutrino-trapping which occurs in the dense supernova core where a
neutron star is formed.

This work is a continuation of our previous work concerning the
EOS of hot $\beta$-stable nuclear matter in cases where neutrinos
have left the system \cite{Moustakidis-08-2}. More specifically,
in order to study the properties and the EOS of hot nuclear
matter, a momentum-dependent effective interaction model (MDIM)
has been applied, one which is able to reproduce the results of
more microscopic calculations of dense matter at zero temperature
and which can be extended to finite temperature
\cite{Prakash-97,Moustakidis-08-2,Moustakidis-07,Moustakidis-08-1}.
 The main incentive for the present study is the fact that only
few calculations of the equation of state of the supernova matter
at high densities are available, although  at lower densities
($n<n_0$) (where $n_0$ is the saturation density) reliable results
are already available.


\section{The model}

The model we use here, which has already been presented and
analyzed in our previous papers
\cite{Moustakidis-08-2,Moustakidis-07,Moustakidis-08-1,Moustakidis-07-a},
is designed to reproduce the results of the microscopic
calculations of both nuclear and neutron-rich matter at zero
temperature and can be extended to finite temperature
\cite{Prakash-97,Li-08}. The energy density of the asymmetric
nuclear matter (ANM) is given by the relation
\begin{equation}
\epsilon(n_n,n_p,T)=\epsilon_{kin}^{n}(n_n,T)+\epsilon_{kin}^{p}(n_p,T)+
V_{int}(n_n,n_p,T), \label{E-D-1}
\end{equation}
where $n_n$ ($n_p$) is the neutron (proton) density and the total
baryon density is $n=n_n+n_p$. The contribution of the kinetic
parts are
\begin{equation}
\epsilon_{kin}^n(n_n,T)+\epsilon_{kin}^p(n_p,T)=2 \int \frac{d^3
k}{(2 \pi)^3}\frac{\hbar^2 k^2}{2m} \left(f_n(n_n,k,T)+
f_p(n_p,k,T) \right), \label{E-K-D-1}
\end{equation}
where $f_{\tau}$, (for $\tau=n,p$) is the Fermi-Dirac distribution
function.

Including the effect of finite-range forces between nucleons, the
potential contribution is parameterized as follows
\cite{Prakash-97}
\begin{eqnarray}
V_{int}(n_n,n_p,T)&=&\frac{1}{3}An_0\left[\frac{3}{2}-(\frac{1}{2}+x_0)I^2\right]u^2
+\frac{\frac{2}{3}Bn_0\left[\frac{3}{2}-(\frac{1}{2}+x_3)I^2\right]u^{\sigma+1}}
{1+\frac{2}{3}B'\left[\frac{3}{2}-(\frac{1}{2}+x_3)I^2\right]u^{\sigma-1}}
\nonumber \\ &+& u \sum_{i=1,2}\left[C_i \left({\cal J}_n^i+{\cal
J}_p^i\right)  + \frac{(C_i-8Z_i)}{5}I\left({\cal J}^i_n-{\cal
J}_p^i\right)\right], \label{V-all}
\end{eqnarray}
where
\begin{equation}
{\cal J}_{\tau}^i= \ 2 \int \frac{d^3k}{(2\pi)^3}
g(k,\Lambda_i)f_{\tau}(n_{\tau},k,T). \label{J-tau}
\end{equation}

In Eq.~(\ref{V-all}), $I$ is the asymmetry parameter
($I=(n_n-n_p)/n$) and $u=n/n_0$, with $n_0$ denoting the
equilibrium symmetric nuclear matter density, $n_0=0.16$
fm$^{-3}$. The asymmetry parameter $I$ is related to the proton
fraction $Y_p$ by the equation $I=(1-2Y_p)$.  The parameters $A$,
$B$, $\sigma$, $C_1$, $C_2$ and $B'$ which appear in the
description of symmetric nuclear matter are determined in order
that $E(n=n_0)-mc^2=-16$ {\rm MeV}, $n_0=0.16$ fm$^{-3}$, and the
incompressibility are $K=240$ {\rm MeV}.  The additional
parameters $x_0$, $x_3$, $Z_1$, and $Z_2$ used to determine the
properties of asymmetric nuclear matter are treated as parameters
constrained by empirical knowledge \cite{Prakash-97}.

The function, $g(k,\Lambda_i)$, suitably chosen to simulate finite
range effects, has the following form
\begin{equation}
 g(k,\Lambda_i)=\left[1+\left(\frac{k}{\Lambda_{i}}\right)^2
\right]^{-1}, \label{g-1} \end{equation}
where the finite range parameters are $\Lambda_1=1.5 k_F^{0}$ and
$\Lambda_2=3 k_F^{0}$ and $k_F^0$ is the Fermi momentum at the
saturation point $n_0$.

The energy density of asymmetric nuclear matter at density $n$ and
temperature $T$, in a good approximation, is expressed as
\begin{equation}
\epsilon(n,T,I)=\epsilon(n,T,I=0)+\epsilon_{sym}(n,T,I),
\label{e-asm-1}
\end{equation}
where
\begin{equation}
\epsilon_{sym}(n,T,I)=nI^2 E_{sym}^{tot}(n,T) =n I^2
\left(E_{sym}^{kin}(n,T)+E_{sym}^{int}(n,T)\right).
\label{e-sym-1}
\end{equation}
In Eq.~(\ref{e-sym-1}) the nuclear symmetry energy
$E_{sym}^{tot}(n,T)$ is separated into two parts corresponding to
the kinetic contribution $E_{sym}^{kin}(n,T)$ and the interaction
contribution $E_{sym}^{int}(n,T)$.

From Eqs.~(\ref{e-asm-1}) and (\ref{e-sym-1}) and setting $I=1$,
we find that the nuclear symmetry energy $E_{sym}^{tot}(n,T)$ is
given by
\begin{equation}
E_{sym}^{tot}(n,T)=\frac{1}{n}\left(\epsilon(n,T,I=1)-\epsilon(n,T,I=0)
\right). \label{Esym-d-1}
\end{equation}
Thus, from Eq.~(\ref{Esym-d-1}) and by a suitable choice of the
parameters $x_0$, $x_3$, $Z_1$ and $Z_2$, we can obtain different
forms for the density dependence of the symmetry energy
$E_{sym}^{tot}(n,T)$.

In a very recent work, \cite{Sammarruca-08} the authors carried
out a systematic analysis of the nuclear symmetry energy in the
formalism of the relativistic Dirac-Brueckner-Hartree-Fock
approach. In this case $E_{sym}(u)$ is obtained with the simple
parametrization $E_{sym}(u)=C u^{\gamma}$ with $\gamma=0.7-1.0$
and $C\approx 32$ MeV.  The authors concluded that a value of
$\gamma$ close to $0.8$ gives a reasonable description of their
predictions although the use of different functions in different
density regions may be best for an optimal fit
\cite{Sammarruca-08}. The results of Ref.~\cite{Sammarruca-08} are
well reproduced by parameterizing the nuclear symmetry energy
according to the following formula
\begin{equation}
E_{sym}^{tot}(n,T=0)= 13 u^{2/3}+17 F(u),\label{Esym-3}
\end{equation}
where the first term of the right part of Eq.~(\ref{Esym-3})
corresponds to the contribution of the kinetic energy and the
second term to the contribution of the interaction energy.

For the function $F(u)$, which parameterizes the interaction part
of the symmetry energy, we apply the following form
\begin{equation}
F(u)=u. \label{Fu-form}
\end{equation}
The parameters $x_0$, $x_3$, $Z_1$ and $Z_2$ are chosen so that
Eq.~(\ref{Esym-d-1}), for $T=0$ reproduces the results of
Eq.~(\ref{Esym-3}) for  the function $F(u)=u$.

\subsection{Thermodynamic description of hot nuclear matter}
\begin{figure}[p]
\centering
\includegraphics[width=59mm]{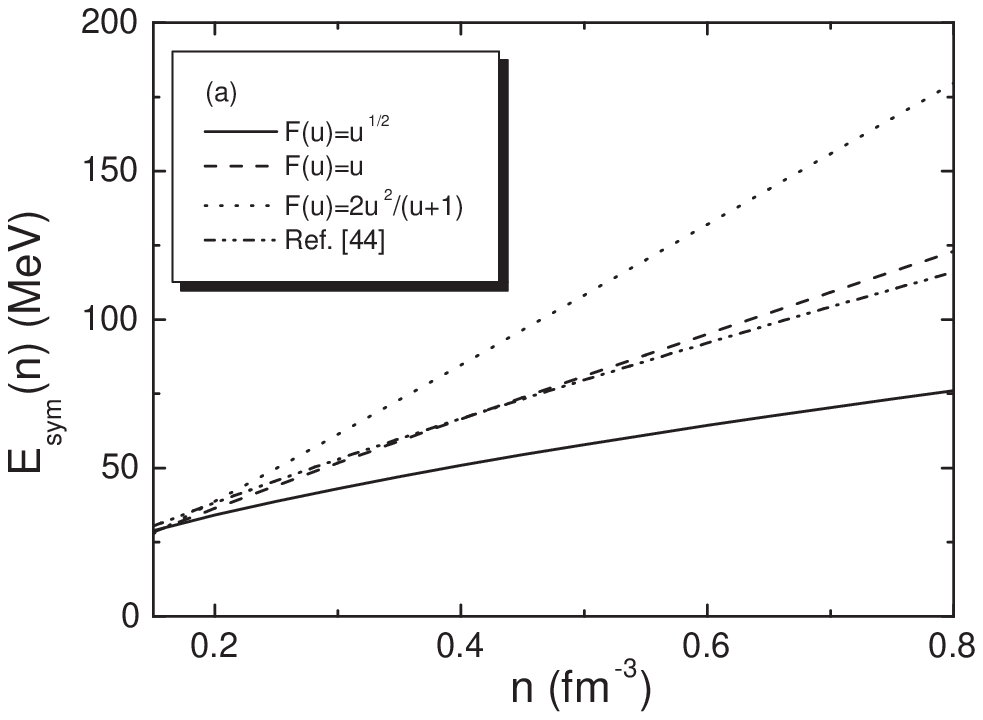}
\includegraphics[width=57mm]{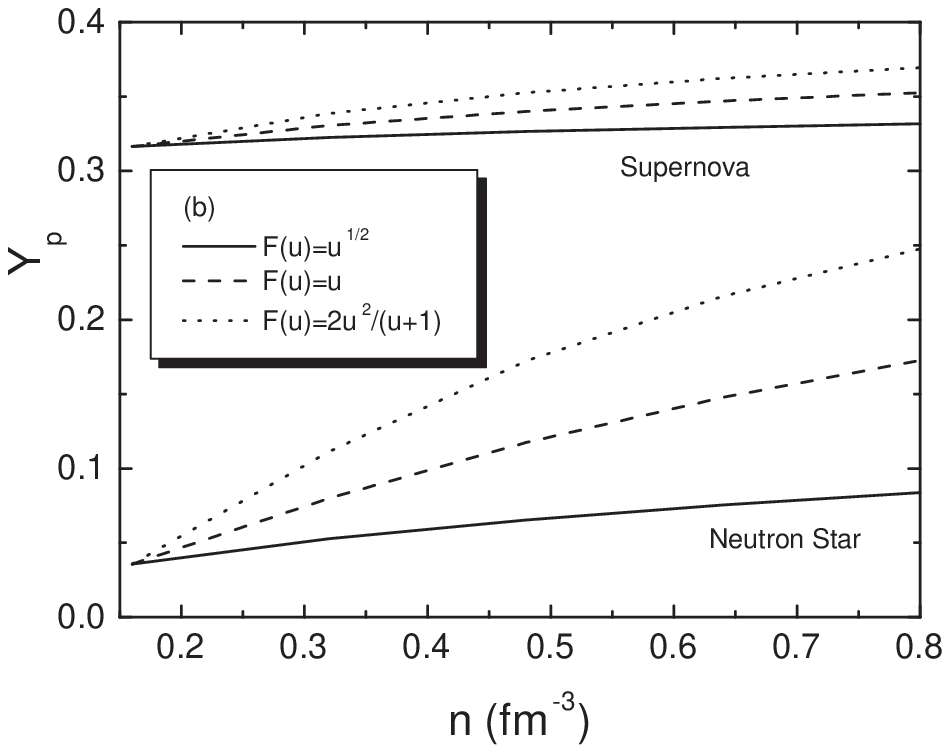}
\caption{ a) The nuclear symmetry energy for three different
parametrization of the interaction part with the results of
reference~\cite{Sammarruca-08} (see text for more details) and b)
the proton fraction $Y_p$ versus baryon density for cold neutron
star matter (down curves) and supernova matter (up curves) for the
three different parametrization of the nuclear symmetry energy. }
\label{}
\end{figure}

In order to study the properties of nuclear matter at finite
temperature, we need to introduce the Helmholtz free energy $F$
which is written as \cite{Fetter-03}

\begin{equation}
F(n,T,I)=E(n,T,I)-TS(n,T,I). \label{Free-1}
\end{equation}
In Eq.~(\ref{Free-1}),  $E$ is the internal energy per particle,
$E=\epsilon/n$, and $S$ is the entropy per particle, $S=s/n$. From
Eq.~(\ref{Free-1}) it is also concluded that for $T=0$, the free
energy $F$ and the internal energy $E$ coincide.

The entropy density $s$ has the same functional form as that of a
non-interacting gas system, given by the equation
\begin{equation}
s_{\tau}(n,T,I)=-2\int \frac{d^3k}{(2\pi)^3}\left[f_{\tau} \ln
f_{\tau}+(1-f_{\tau}) \ln(1-f_{\tau})\right],  \label{s-den-1}
\end{equation}
while the pressure and the chemical potentials defined as follows
\cite{Fetter-03}
\begin{equation}
P=n^2\left(\frac{\partial (\epsilon/n)}{\partial
n}\right)_{S,N_i}, \qquad \qquad \qquad
\mu_{i}=\left(\frac{\partial \epsilon}{\partial
n_i}\right)_{S,V,n_{j\neq i}}. \label{P-m-E}
\end{equation}

At this point we shall examine the properties and the EOS of
nuclear matter by considering an isothermal process. In this case,
the pressure and the chemical potentials are related to  the
derivative of the total free energy density $f=F n$. More
specifically, they are defined as follows
\begin{equation}
P=n^2\left(\frac{\partial (f/n)}{\partial n}\right)_{T,N_i},
\qquad \qquad \mu_{i}=\left(\frac{\partial f}{\partial
n_i}\right)_{T,V,n_{j\neq i}}. \qquad \qquad \label{P-m-F}
\end{equation}



%
It is easy to demonstrate by applying Eq.~(\ref{P-m-F}) that (see
for a proof \cite{Prakash-94} as well as \cite{Burgio-07})
\begin{eqnarray}
\mu_n&=&F+u\left(\frac{\partial F}{\partial
u}\right)_{Y_p,T}-Y_p\left(\frac{\partial F}{\partial
Y_p}\right)_{n,T}, \nonumber
\\ \mu_p&=&\mu_n+\left(\frac{\partial F}{\partial
Y_p}\right)_{n,T}, \nonumber \\
\hat{\mu}&=&\mu_n-\mu_p=-\left(\frac{\partial F}{\partial
Y_p}\right)_{n,T}.
 \label{mu-p-n}
\end{eqnarray}

We can define the symmetry free energy per particle $F_{sym}(n,T)$
by the following parabolic approximation (see also
\cite{Burgio-07,Xu-07-3})
\begin{eqnarray}
F(n,T,I)&=&F(n,T,I=0)+I^2F_{sym}(n,T)\nonumber \\
&=&
F(n,T,I=0)+(1-2Y_p)^2F_{sym}(n,T),\label{Free-Parabolic}
\end{eqnarray}
where
\begin{equation} F_{sym}(n,T)= F(n,T,I=1)-F(n,T,I=0).
\label{Free-asym}
\end{equation}
It is worth noting that the above approximation is not valid from
the beginning, but one needs to check the validity of the
parabolic law in the present model before using it. In
Ref.~\cite{Moustakidis-08-2} we have proved the validity of the
approximation (\ref{Free-Parabolic}).

Now, by applying Eq.~(\ref{Free-Parabolic}) in Eq.~(\ref{mu-p-n}),
we obtain the  key relation
\begin{equation}
\hat{\mu}=\mu_n-\mu_p=4(1-2Y_p)F_{sym}(n,T). \label{mhat-2}
\end{equation}
The above equation is similar to that obtained for cold nuclear
matter by replacing $E_{sym}(n)$ with $F_{sym}(n,T)$.

\subsection{$\beta$-equilibrium in hot proto-neutron star and supernova}

Stable high density nuclear matter must be in chemical equilibrium
with all type of reactions, including the weak interactions in
which $\beta$ decay and electron capture take place simultaneously
\begin{equation}
n \longrightarrow p+e^{-}+\bar{\nu}_e, \qquad \qquad p +e^{-}
\longrightarrow n+ \nu_e. \label{b-reaction}
\end{equation}
Both types of reactions change the electron per nucleon fraction,
$Y_e$ and thus affect the equation of state. In a previous study,
we assumed that neutrinos generated in these reactions left the
system \cite{Moustakidis-08-2}. The absence of neutrino-trapping
has a dramatic effect on the equation of state and is the main
cause of  a significant reduction in the values of the proton
fraction $Y_p$ \cite{Takatsuka-94,Takatsuka-96}.
The equation of state of hot nuclear matter in $\beta$-equilibrium
(considering that it consists of neutrons, protons, electrons and
neutrinos) can be obtained by calculating the total energy density
$\epsilon_{tot}$ as well as the total pressure $P_{tot}$. The
total energy density is given by
\begin{equation}
\epsilon_{tot}(n,T,I)=\epsilon_b(n,T,I)+\sum_{l=e,\nu_e}\epsilon_l(n,T,I),
\label{e-de-1}
\end{equation}
where $\epsilon_b(n,T,I)$ and $\epsilon_l(n,T,I)$ are the
contributions of baryons and leptons respectively. The total
pressure is
\begin{equation}
P_{tot}(n,T,I)=P_b(n,T,I)+\sum_{l=e,\nu_e}P_l(n,T,I), \label{Pr-1}
\end{equation}
where $P_b(n,T,I)$ is the contribution of the baryons i.e.
\begin{equation}
P_b(n,T,I)=T\sum_{\tau=p,n}s_{\tau}(n,T,I)+\sum_{\tau=n,p}n_{\tau}\mu_{\tau}(n,T,I)-\epsilon_b(n,T,I),
\label{Pr-2}
\end{equation}
while $P_l(n,T,I)$ is the contribution of the leptons. From
Eqs.~(\ref{e-de-1}) and (\ref{Pr-1}) we can construct the
isothermal curves for energy and pressure and finally derive the
isothermal behavior of the equation of state of hot nuclear matter
under $\beta$-equilibrium.

\section{Results and Discussions}

\begin{figure}[p]
\centering
\includegraphics[width=38.5mm]{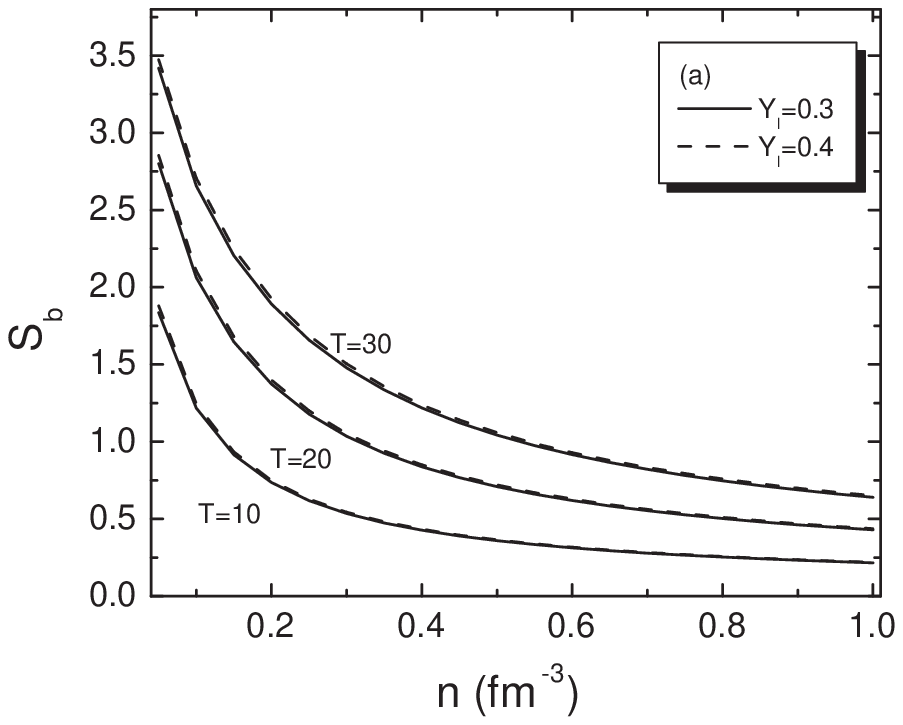}
\includegraphics[width=38.4mm]{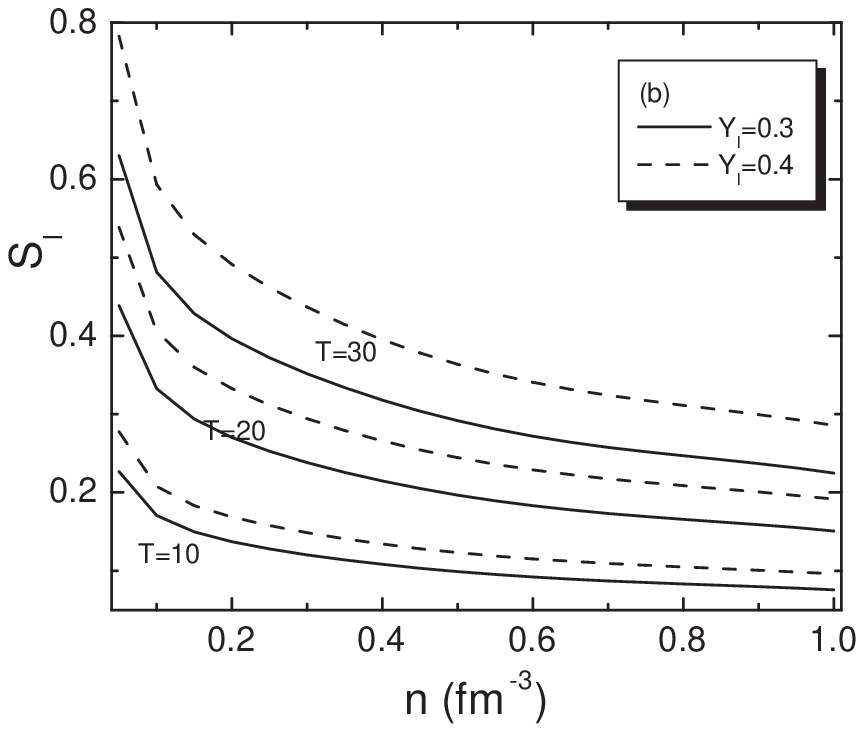}
\includegraphics[width=38.4mm]{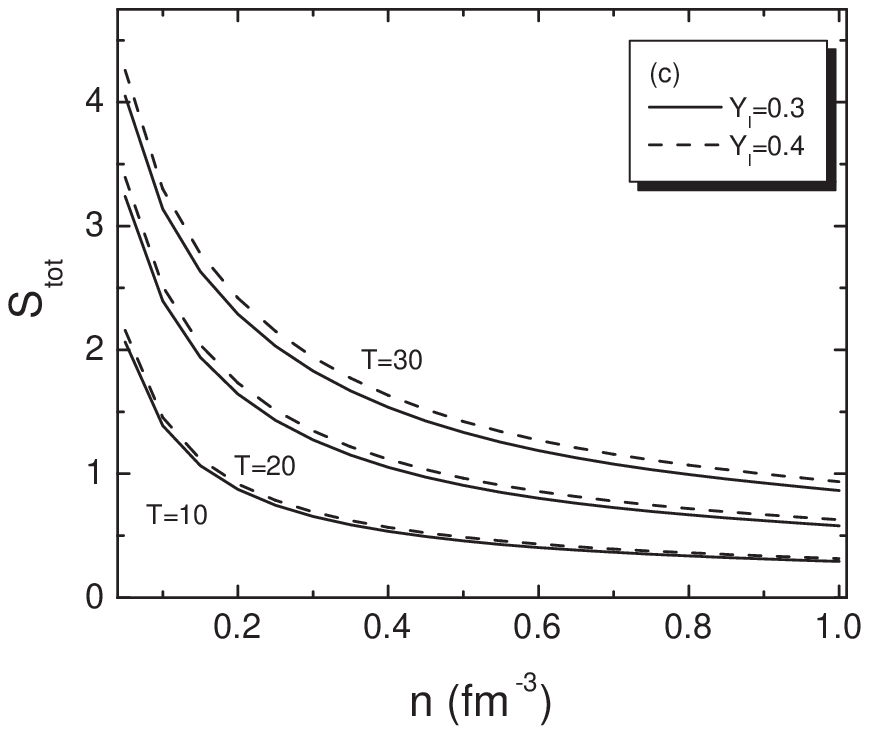}
\caption{Contribution to the total entropy per particle of a)
baryons ($S_b$), b) leptons ($S_l$) and c) the total entropy
($S_{tot}$) versus the baryon density for various values of T for
total lepton fraction $Y_l=0.3$ and $Y_l=0.4$. } \label{}
\end{figure}
\begin{figure}[p]
\centering
\includegraphics[width=59mm]{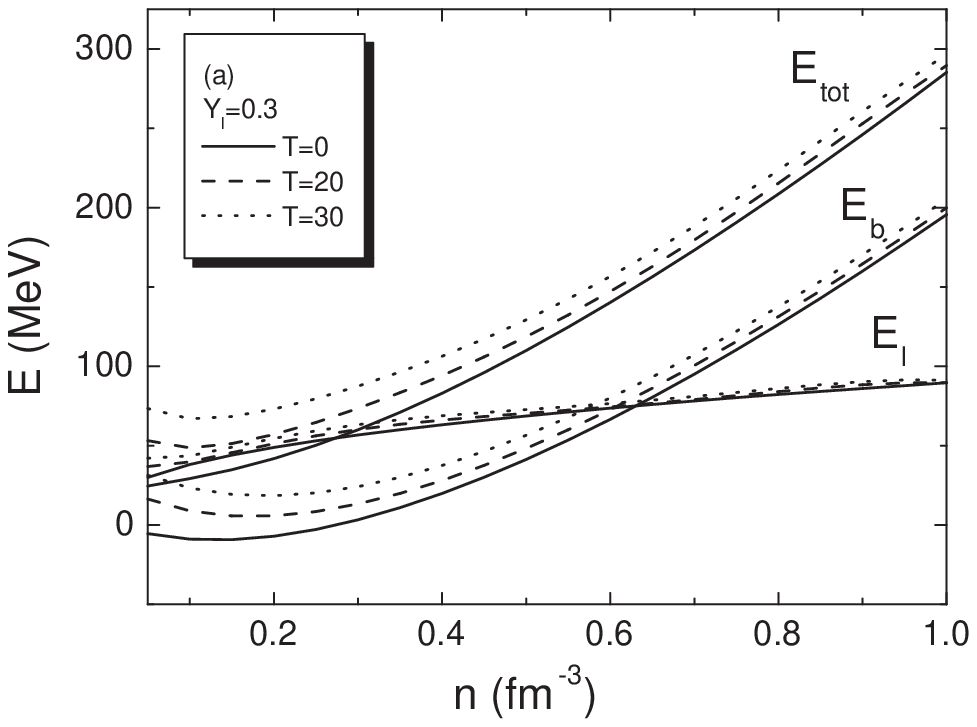}
\includegraphics[width=57mm]{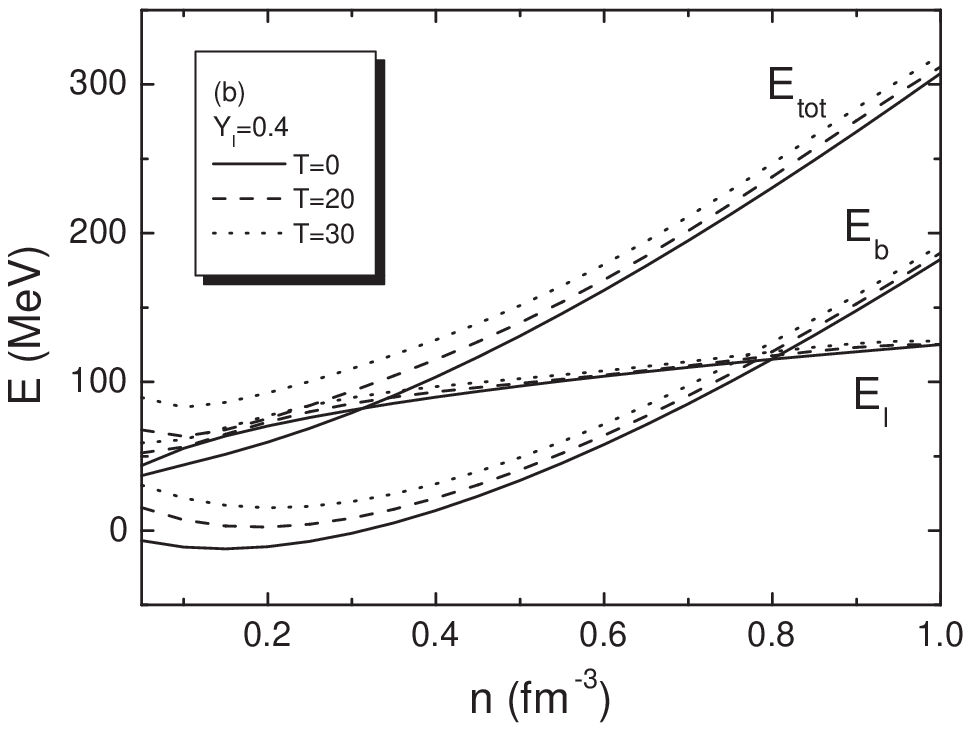}
\caption{Contribution to the total energy per particle of baryons
($E_b$), leptons ($E_l$) and  the total energy ($E_{tot}$) versus
the baryon density for various values of T for total lepton
fraction  a) $Y_l=0.3$ and b) $Y_l=0.4$.} \label{}
\end{figure}

We calculate the equation of state of hot asymmetric nuclear
matter by applying a momentum dependent effective interaction
model describing the baryons interaction. We consider that nuclear
matter contains neutrons, protons, electrons and neutrinos under
$\beta$-equilibrium and charge neutrality. The key quantities in
our calculations are the proton fraction $Y_p$ and also the
asymmetry free energy defined in Eq.~(\ref{Free-asym}). It is
worth pointing out that since the supernova explosion itself is a
dynamic phenomenon, the chemical composition of matter changes
according to the evolution of the star all the time
\cite{Sumiyoshi-94}. During supernova explosion, the chemical
composition of matter reaches equilibrium not in the whole star
 but locally. In our present work we assume matter in the
chemical equilibrium for simplicity in order to analyze the
properties of hot neutron star and supernova matter.

Following the discussion of Takatsuka et al. \cite{Takatsuka-94},
we attempt to extend the discussion concerning the dependence of
equilibrium  fraction $Y_e$($= Y_p$) on the baryon density as well
as on the nuclear symmetry energy. We ignore the temperature
effect to clarify the situation. Actually, the situation does not
change by including finite temperature effects. The energy per
baryon of supernova matter $E_{SM}$ and cold neutron star matter
$E_{NS}$ are expressed as function of $n$ and $Y_p$ (see also
ref.~\cite{Takatsuka-94}) as
\begin{eqnarray}
E_{SM}(n,Y_p)&=&E_b(n,Y_p)+E_e(n,Y_p)+E_{\nu_e}(n,Y_p) \\
&=&E_b(n,Y_p=0.5)+E_{sym}(n)(1-2Y_p)^2 \nonumber \\
&+&253.6
u^{1/3}Y_p^{4/3}+319.516u^{1/3}(Y_l-Y_p)^{4/3},\nonumber \\
E_{NS}(n,Y_p)&=&E_b(n,Y_p)+E_e(n,Y_p) \\
&=&E_b(n,Y_p=0.5)+E_{sym}(n)(1-2Y_p)^2+253.6
u^{1/3}Y_p^{4/3}.\nonumber \label{E-NSM}
\end{eqnarray}
$E_{sym}(n)$ is plotted in Fig. 1(a) for the three different
parametrizations. In the same figure we have included recent
results provided in reference~\cite{Sammarruca-08} achieved by
performing microscopic calculations in asymmetric nuclear matter.
In this case $E_{sym}(n)$ is obtained with the simple
parametrization
\[E_{sym}(u)=C u^{\gamma},\]
with $\gamma=0.8$ and $C=32$ MeV. It is obvious that the results
of the above parametrization, correspond very well with the
parametrization $F(u)=u$ which is proposed here.

The equilibrium proton fraction $Y_p$ is calculated by solving the
equation $\partial E_{SM,NS}/\partial Y_p=0$ for various values of
the density $n$, $E_{sym}(n)$ and $Y_l=0.4$ for supernova matter
(see Fig.~1(b)). In the case of cold neutron star matter, $Y_p$
depends strongly on both  the baryon density and the values of the
$E_{sym}(n)$. This is not the case for supernova matter where the
effect of nuclear symmetry energy in determining $Y_p$ is less
important than in cold neutron star matter. In addition, $Y_p$,
for a fixed parametrization of $F(u)$ is almost constant with
respect to $n$.
\begin{figure}[p]
\centering
\includegraphics[width=57mm]{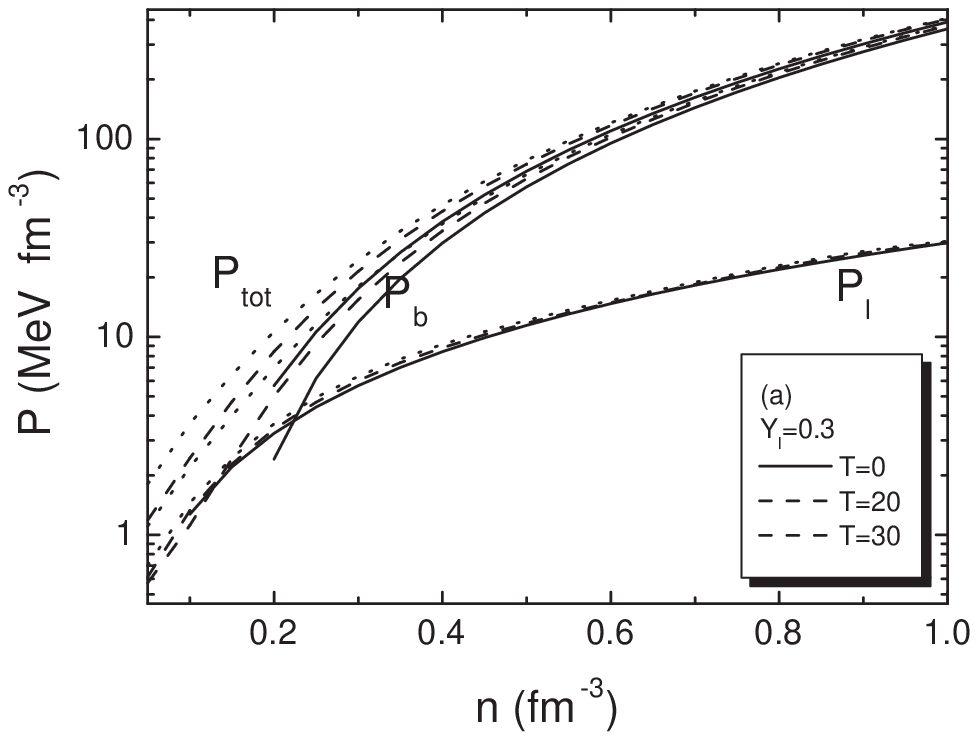}
\includegraphics[width=57mm]{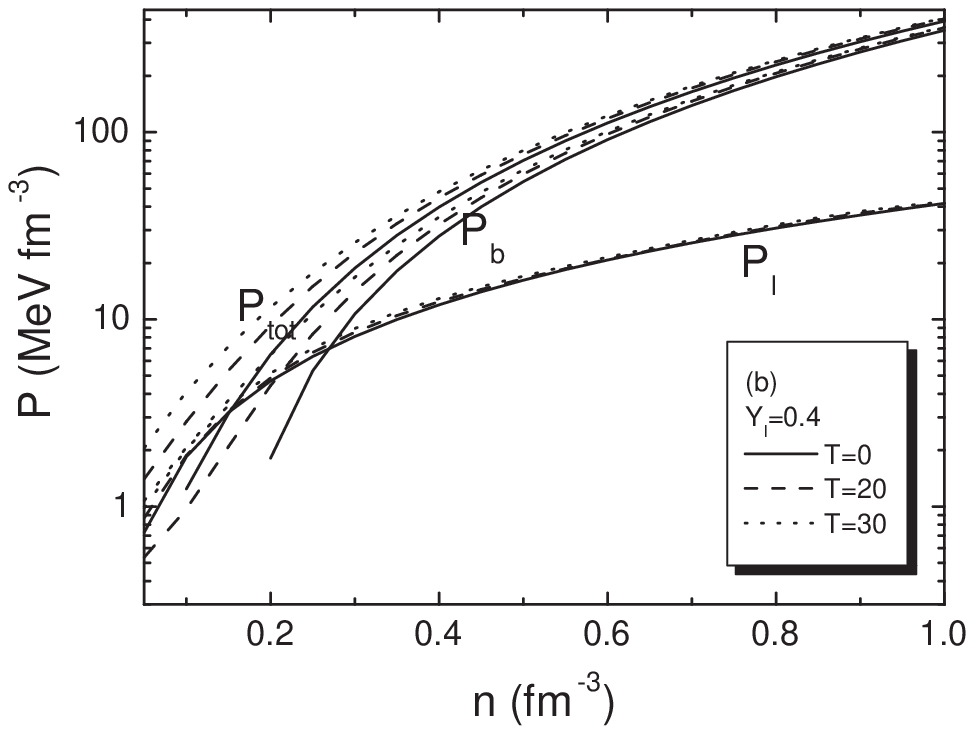}
\caption{The pressure of baryons ($P_b$), leptons ($P_l$) and the
total pressure ($P_{tot}$) versus the baryon density for various
values of T for the cases a) $Y_l=0.3$ and b) $Y_l=0.4$. }
\label{}
\end{figure}

In Fig.~2 we plot the contribution of the baryons $S_b$, leptons
$S_l$ and the total $S_{tot}$ to the entropy per baryon. In all
cases, $S$ is a decreasing function of the baryon density $n$.
Temperature affects appreciably both baryon and lepton
contribution. It should be noted that the contribution of baryons
$S_b$ may be written as $S_b=S_{kin}+S_{int}$, where the term
$S_{kin}$  originates from  the temperature effect on the kinetic
part of the energy density and $S_{int}$ reflects thermal effects
on the potential energy density.
\begin{figure}[p]
\centering
\includegraphics[width=55mm]{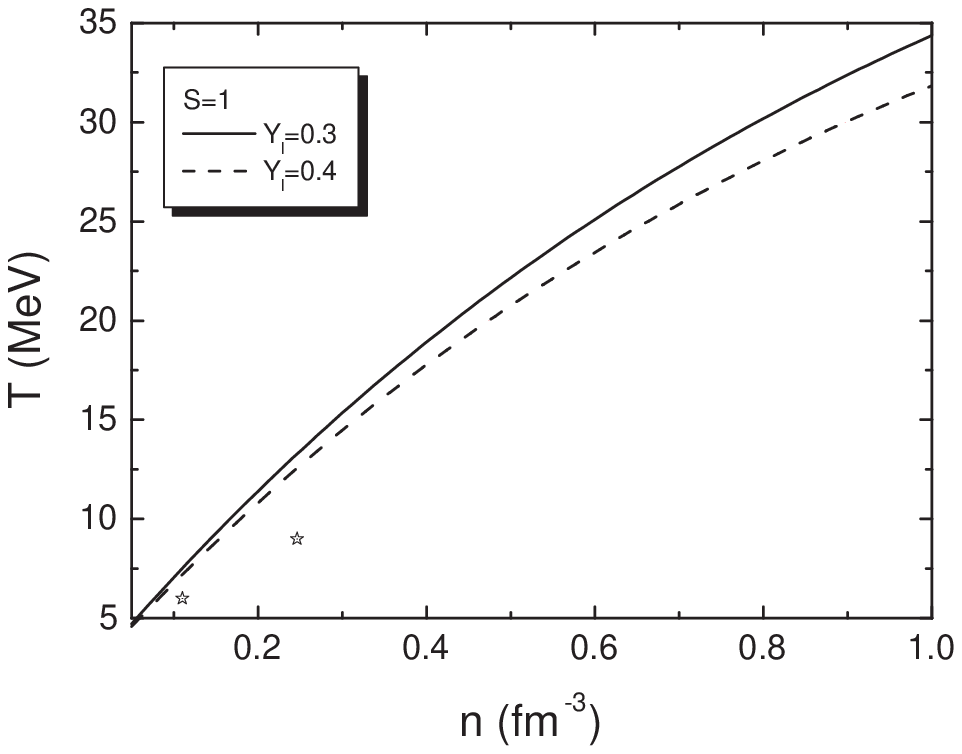}
\caption{Temperature $T$-density $n$ relation with $Y_l=0.3$
(solid line) and $Y_l=0.4$ (dashed line) for $S=1$. Stars denote
the results for the case with  $S=1$ and $Y_l=0.4$, extracted from
the results by Lattimer et al.~\cite{Lattimer-85}. } \label{}
\end{figure}
\begin{figure}[p]
\centering
\includegraphics[width=55mm]{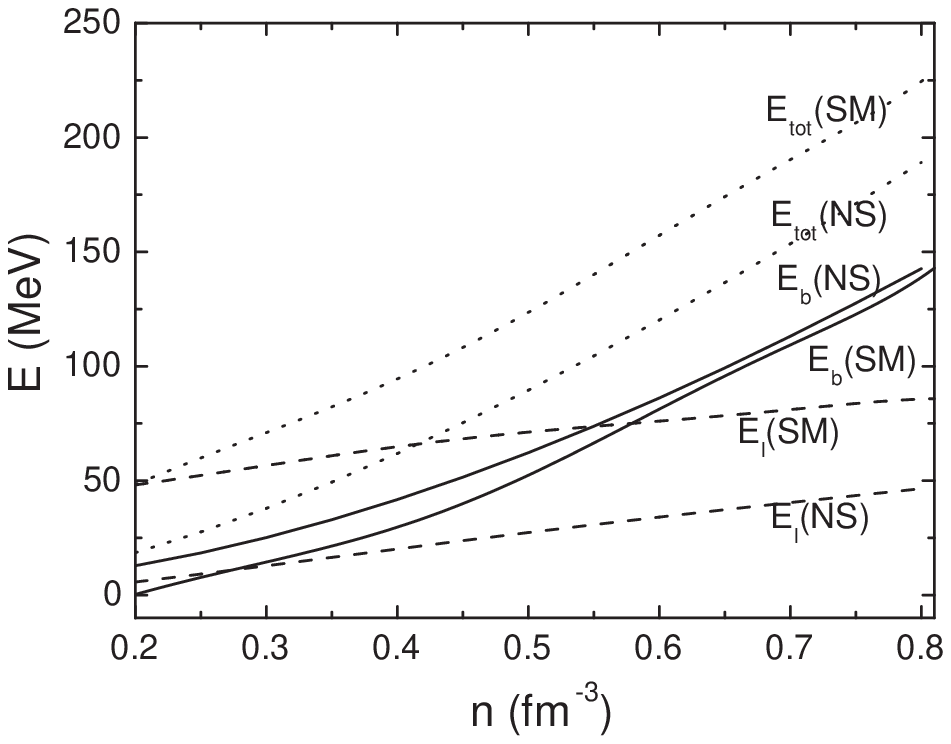}
\caption{Internal energy per baryon versus density $n$ for dense
supernova matter (SM) in comparison with that of cold neutron star
matter (NS) by applying the same model. The case of supernova
matter corresponds to $S=1$ and $Y_l=0.3$. The contribution of
each species is plotted separately.} \label{}
\end{figure}

In Fig.~3, we display the contribution to internal energy $E$ from
baryons $E_b$ and leptons $E_l$  for $Y_l=0.3$ and $Y_l=0.4$ and
for various values of $T$. The most striking aspect is that the
lepton energy, $E_l=E_e+E_{\nu_e}$, dominates in the internal
energy of the matter up to $n\sim 0.6$ fm$^{-3}$ (for $Y_l=0.3$)
and $n\sim 0.8$ fm$^{-3}$ (for $Y_l=0.4$). This is a
characteristic of the supernova matter and is in remarkable
contrast with the situation of cold neutron star matter
\cite{Takatsuka-94}. The contribution from baryon $E_b$ gets
larger with the increase of $n$ and is comparable with $E_l$ for
high values of $n$.

The contributions of baryon and leptons on the total pressure are
presented in Fig.~4. In contrast to the situation of the internal
energy, the nuclear part contribution plays a more important role
compared with the lepton part. The lepton pressure $P_l$ is
comparable to baryon pressure $P_b$ up to $n\sim 0.2$ fm$^{-3}$,
but for higher values of $n$ it is significantly small.

As pointed out by Bethe et al. \cite{Bethe-79}, the crucial
feature in determining the evaluation of a collapsing
pre-supernova core is that the entropy per particle is very low,
of the order of unity (in units of the Boltzmann constant $k_B$),
and nearly constant during all the stages of the collapse up to
the shock wave formation. Therefore, the collapse is an adiabatic
process of a highly ordered system. So, since the supernova matter
is characterized by a constant entropy and constant lepton
fraction, we shall also discuss the properties under this
condition. This can be done by converting the results for
isothermal case ($T$=const) into those for adiabatic case
($S$=const) in terms of the $T-n$ relation constrained by a
constant entropy.

The $T=T(n)$ relation is constructed by $\{T,n\}$ values to
satisfy $S(n,T)$=const in an $S-n$ diagram. Fig.~5 shows the
results for $Y_l=0.3$ and $Y_l=0.4$ for $S=1$. Temperature is an
increasing function of $n$. Furthermore, for the same density, the
temperature is higher for lower values of $Y_l$. The values of $T$
for various values of $n$ are derived, for the two cases, with the
least-squares fit method and found to take the general form
\[T(n)=an^{b},\]
where $a=35.412$, $b=0.70481$ for $Y_l=0.3$ and $a=32.35706$,
$b=0.67694$ for $Y_l=0.4$. The results of this study are in
 very good agreement with those of Takatsuka et al.
\cite{Takatsuka-94}. The stars at lower density denote the
$\{T,n\}$ values for $S=1$ and $Y_l=0.4$ which are derived from
Lattimer et al. \cite{Lattimer-85}. It is concluded that the
temperature increases considerably when moving from the outer part
of the star to the center in order to maintain a constant value of
the entropy per baryon.

Finally, in Fig.~6 we compare the EOS's between supernova matter
and cold neutron star matter. The case for supernova matter
corresponds to $S=1$ and $Y_l=0.3$. It is thus clear that the
internal energy $E_{tot}$ of supernova matter (SM) is remarkably
larger than that of neutron star matter (NS). As far as the
nucleon part $E_b$ is concerned, the $E_b$ in SM is slightly lower
than that in NS due to the large energy gain in symmetry energy
(see also \cite{Takatsuka-94}). However, the lepton contribution
on the internal energy $E_l$ is remarkably larger in SN matter
compared to NS matter due to the effect of a large lepton
fraction, that is, a large kinetic energy of abundant leptons.
High temperature also contributes to the stiffening, but it is
less effective than the high lepton fractions.

\section{Summary}
The evaluation of the equation of state of hot nuclear matter is a
major challenge for nuclear physics and astrophysics. EOS is the
basic ingredient necessary for studying the supernova explosion as
well as for determining the properties of  hot neutron stars.  The
motive for the present work has been to apply a momentum-dependent
interaction model for the study of the hot nuclear matter EOS
under $\beta$-equilibrium.  Special attention has been dedicated
to the study of the contribution of the components of
$\beta$-stable nuclear matter on the entropy per particle, a
quantity of great interest in the study of structure and collapse
of supernova.  The above EOS can be applied to the evaluation of
the gross properties of hot neutron stars i.e. mass and radius.

\section*{Acknowledgments}
The author  would like to thank Professor Tatsuyauki Takatsuka for
valuable comments and correspondence.


\end{document}